\documentclass[aps,twocolumn,superscriptaddress,showpacs]{revtex4}
\usepackage{graphicx}
\begin{document}
\newcommand{\tdna}{t^{\dagger}_{n\alpha}}
\newcommand{\tna}{t_{n\alpha}}
\newcommand{\tdnoa}{t^{\dagger}_{n+1,\alpha}}
\newcommand{\tnoa}{t_{n+1,\alpha}}
\newcommand{\tdng}{t^{\dagger}_{n\gamma}}
\newcommand{\tng}{t_{n\gamma}}
\newcommand{\tdnod}{t^{\dagger}_{n+1,\delta}}
\newcommand{\tnod}{t_{n+1,\delta}}
\newcommand{\tdag}{t^{\dagger}}
\newcommand{\al}{\alpha}
\newcommand{\ca}{\gamma}
\newcommand{\de}{\delta}
\newcommand{\taud}{\tau^{\dagger}}
\newcommand{\bfdelta}{ \mbox{\boldmath $\delta$} }
\newcommand{\sbfdelta}{\mbox{{\scriptsize\boldmath $\delta$}}}
\newcommand{\be}{\begin{equation}}
\newcommand{\ee}{\end{equation}}
\newcommand{\bea}{\begin{eqnarray}}
\newcommand{\eea}{\end{eqnarray}}
\newcommand{\half}{ \frac{1}{2} }   

\title{ Critical Behaviour of One-particle Spectral Weights
 in the Transverse
Ising Model}
\author{C. J. Hamer}
\affiliation{School of Physics, University of New South Wales,
Sydney NSW 2052, Australia}
\author{J. Oitmaa}
\affiliation{School of Physics, University of New South Wales,
Sydney NSW 2052, Australia}
\author{Weihong Zheng}
\affiliation{School of Physics, University of New South Wales,
Sydney NSW 2052, Australia}
\author{Ross H. McKenzie}
\affiliation{Department of Physics, University of Queensland,
Brisbane, Qld 4072, Australia}

\date{\today}
\begin{abstract}
We investigate the critical behaviour of the spectral weight of a single
quasiparticle, one of the key observables in experiment, for the
particular case of the transverse Ising model.
Series expansions are calculated for the
linear
chain and the
square and simple cubic lattices. For the chain model,
a conjectured exact result is discovered.
For the square and simple cubic lattices, series analyses are
used to estimate the critical exponents. The results agree
with the general predictions of Sachdev.
\end{abstract}
\maketitle

There is currently much interest in quantum phase transitions in strongly
correlated magnetic and/or electronic systems in condensed matter
physics. The physics in the vicinity of such quantum critical points can
be described via general scaling theories, independent of any particular
microscopic model. Sachdev \cite{sachdev1999} has developed a scaling
theory for the spectral weight (defined below), a quantity which can be
measured experimentally, and thus provides a bridge between model
calculations and real materials. Our goal in this letter is to verify
Sachdev's predictions for a particular quantum spin model, the Ising
model in a transverse field. To the best of our knowledge, this is the
first such microscopic verification of the scaling form.

A fundamental quantity to describe the dynamical properties of quantum
magnets is the dynamical structure factor \cite{marshall1971}
\begin{eqnarray}
S^{\alpha\beta}({\bf k},\omega) & = & \frac{1}{2\pi N}\sum_{i,j}
\int_{-\infty}^{\infty}dt  e^{i(\omega t +{\bf k \cdot (r_i - r_j)) }}
\nonumber \\
 & & \times <S^{\alpha}_j(t)S^{\beta}_i(0)>
\label{eq1}
\end{eqnarray}
where the notation is standard, and the angular bracket denotes a
thermal average or, at zero temperature, a ground state expectation
value. As is well known, $S({\bf k},\omega)$  can be related directly to
the inelastic neutron scattering intensity, allowing a comparison
between experiments on real materials and theoretical model
calculations.

Integrating (\ref{eq1}) over energy, we obtain the integrated or static structure
factor,
\begin{eqnarray}
S^{\alpha\beta}({\bf k}) & = & \frac{1}{N}\sum_{i,j}
 e^{i{\bf k \cdot (r_i - r_j) }}
 <S^{\alpha}_jS^{\beta}_i>
\label{eq1a}
\end{eqnarray}
which is just the Fourier transform of the spin-spin correlation
function.

At low temperatures the major contribution to $S({\bf k},\omega)$ will
often come from low energy quasiparticle excitations. It is then useful
to write (\ref{eq1}) in the `spectral form'
(valid for a discrete spectrum)
\begin{equation}
S^{\alpha\beta}({\bf k},\omega) = \sum_{\Lambda} \delta(\omega+E_0-E_{\Lambda})
S^{\alpha\beta}_{\Lambda}({\bf k})
\label{eq2}
\end{equation}
where the sum is over all eigenstates with energies $E_{\Lambda}$, and
the `spectral weights' $S^{\alpha\beta}_{\Lambda}({\bf k})$ are given by
\begin{equation}
S^{\alpha\beta}_{\Lambda}({\bf k}) =
\Omega_{\Lambda}^{\alpha *}({\bf k})
\Omega_{\Lambda}^{\beta}({\bf k})
\label{eq3}
\end{equation}
with
\begin{equation}
\Omega_{\Lambda}^{\beta}({\bf k})
=\frac{1}{\sqrt N} \sum_i <\psi_{\Lambda}|S^{\beta}_i|\psi_0>e^{i{\bf k
\cdot r}_i}.
\label{eq4}
\end{equation}
We consider only the zero temperature case.

Henceforth we consider only the `one-particle' spectral weights
$S_{\rm 1p}({\bf k})$, referring to a state of a single quasiparticle with
momentum ${\bf k}$. Near a quantum phase transition \cite{sachdev1999}
these quantities are expected to show universal scaling properties,
which can be derived from a field theoretic approach, independently of
the details of a particular model. The spectral weight can be written,
up to constant factors, as
\begin{equation}
S_{\rm 1p}({\bf k}) = \frac{A({\bf k})}{E({\bf k})}
\label{eq5}
\end{equation}
where $A({\bf k})$ is the `quasiparticle residue', i.e. the residue of
the Green's function at the quasiparticle pole, and $E({\bf k})$ is
the quasiparticle energy at momentum ${\bf k}$. Near the
critical point, these functions are expected to scale as
\begin{eqnarray}
A({\bf k}) & \sim & (g_c-g)^{\eta \nu},  \nonumber \\
 E({\bf k_c}) & \sim &
(g_c-g)^{\nu}
\label{eq6}
\end{eqnarray}
where $g$ is a coupling or tuning parameter, ${\bf k_c}$ is the critical
momentum at which the energy gap vanishes, and $\eta,\nu$ are the
standard critical exponents. Then $S_{\rm 1p}$ is expected to scale as
\begin{eqnarray}
S_{\rm 1p}({\bf k}) & \sim & (g_c-g)^{\eta \nu}, \hspace{5mm} {\bf k} \neq
{\bf k_c}; \nonumber \\
S_{\rm 1p}({\bf k_c}) & \sim & (g_c-g)^{(\eta-1)\nu}.
\label{eq7}
\end{eqnarray}

Our aim is to confirm and demonstrate this scaling
behaviour for the transverse field Ising
model
\begin{equation}
H = \sum_i (1-\sigma^z_i) - \lambda \sum_{<ij>} \sigma^x_i \sigma^x_j
\label{eq8}
\end{equation}
where $\sigma^{\alpha}_i = 2S^{\alpha}_i$ are Pauli operators and the
second sum is over nearest neighbour pairs. This model has a quantum
phase transition at $\lambda =1$ in one dimension, and at some specific
$\lambda_c$ for each higher-dimensional lattice. Our approach is to
derive series expansions in $\lambda$ for various one-particle spectral
weights
in the `disordered' phase $\lambda < \lambda_c$,
via the linked-cluster techniques described by Oitmaa {\it et al.}
\cite{oitmaa2006},
and
to analyse the series by standard methods to obtain the critical
behaviour.

\begin{flushleft}
{\bf The Linear Chain}
\end{flushleft}

\begin{figure}[!htb]
\begin{center}
  \includegraphics[scale=0.4]{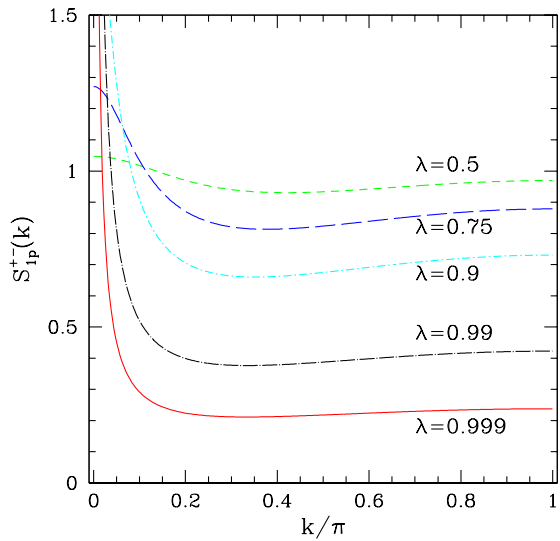}
  \caption{
(Color online) The 1-particle spectral weight $S^{+-}_{1p}(k)$ for the linear chain
model graphed against momentum $k/\pi$, for various couplings $\lambda$.
\label{fig_S1p_pm}
}
\end{center}
\end{figure}
\begin{figure}[!htb]
\begin{center}
  \includegraphics[scale=0.4]{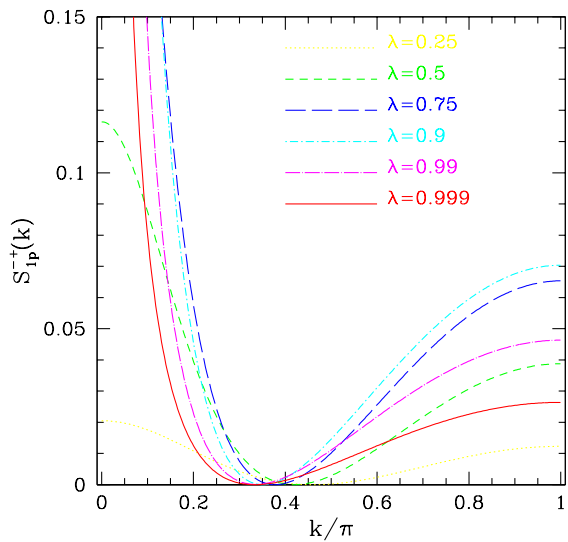}
  \caption{
(Color online) As Fig. \ref{fig_S1p_pm}, for the 1-particle spectral weight
$S^{-+}_{1p}(k)$.
\label{fig_S1p_mp}
}
\end{center}
\end{figure}
\begin{figure}[!htb]
\begin{center}
  \includegraphics[scale=0.42]{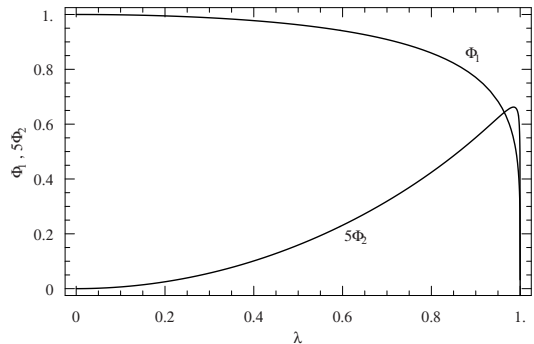}
  \caption{
The integrated 1-particle structure factors $\Phi_1$ and $\Phi_2$
(corresponding to $S^{+-}_{1p}$, $S^{-+}_{1p}$ respectively) graphed as
functions of coupling $\lambda$.
\label{fig_Phi}
}
\end{center}
\end{figure}

The transverse Ising chain model is exactly solvable, and expressions
for the energy spectrum, magnetization, etc. have been given by Pfeuty
\cite{pfeuty1970}.
We have computed series
for the one-particle matrix elements $\Omega_{\rm 1p}^{\pm} (k)$
 for this model
up to order $\lambda^{22}$. The leading terms for $\Omega^-_{1p}(k)$ are given in Ref.
\cite{oitmaa2006}; we can supply further coefficients on request.

While analyzing these series for their critical behaviour, we realized
that they could be represented {\it exactly}, up to the order calculated,
by the closed form expressions:
\be
\Omega^{\pm}_{\rm 1p}(k) = \half (1-\lambda^2)^{1/8}\frac{(1 \mp \Lambda( k))}
{\Lambda( k)^{1/2}}
\label{eq11}
\ee
where $\Lambda( k)$ is just half  the 1-particle energy
$E ({\bf k})$:
\begin{equation}
\Lambda( k) = E ( k) /2 = [1+\lambda^2-2\lambda \cos(k)]^{1/2}.
\label{eq12}
\end{equation}
It is natural to conjecture that this result is exact to all orders.

One can now write down any desired spectral weight
$S_{\rm 1p}^{\alpha\beta} (k)$, e.g.
\bea
S^{+-}_{\rm 1p}(k) &=& \frac{1}{4} (1-\lambda^2)^{1/4}\frac{(1+\Lambda( k))^2}
{\Lambda( k)}, \\
S^{-+}_{\rm 1p}(k) &=& \frac{1}{4} (1-\lambda^2)^{1/4}\frac{(1-\Lambda( k))^2}
{\Lambda( k)}, \\
S^{++}_{\rm 1p}(k) & = & S^{--}_{\rm 1p}(k) = \frac{1}{4} (1-\lambda^2)^{1/4}
\frac{(1-\Lambda^2 (k))}{\Lambda (k)}
\label{eq13}
\eea
These formulae give a very clear demonstration of the critical behaviour
expected theoretically. The quasiparticle residues for $S^{-+}$,
$S^{+-}$ are respectively
\begin{equation}
A ( k) = \frac{1}{2} (1-\lambda^2)^{1/4}(1 {\mp} \Lambda( k))^2
\label{eq14}
\end{equation}
which vanish at the critical point $\lambda = 1$ with exponent $1/4 =
\eta \nu$, as expected for the transverse Ising chain where $\eta =
1/4$, $\nu = 1$. At the critical momentum $k = 0$, the energy gap is
$2(1-\lambda)$, and vanishes with exponent $\nu = 1$.

The behaviour of $S^{+-}_{\rm 1p}(k)$ and $S^{-+}_{\rm 1p}(k) $ for several $\lambda$
is shown in Figs. \ref{fig_S1p_pm}
and \ref{fig_S1p_mp}. We can see that both spectral weights slowly
decrease towards zero as $\lambda \to 1$, except at the point $k=0$
where they both diverge, just as we should expect. Note that  $S^{-+}_{\rm 1p}(k) $
vanishes  at $\lambda=2 \cos( k)$.

The integrated 1-particle structure factors are
\bea
\Phi_1 &=& {1\over \pi} \int_0^{\pi} S^{+-}_{\rm 1p} d k \nonumber \\
       &=& (1-\lambda^2)^{1/4} [ (1+\lambda) E(m)
 \nonumber \\
 & &
  + (1+\lambda)^{-1} K(m) + \pi ]/(2 \pi)  \\
\Phi_2 &=& {1\over \pi} \int_0^{\pi} S^{-+}_{\rm 1p} d k \nonumber \\
       &=& (1-\lambda^2)^{1/4} [ (1+\lambda) E(m)
\nonumber \\
 & &
+ (1+\lambda)^{-1} K(m) - \pi ]/(2 \pi)
\eea
where $m=4 \lambda/(1+\lambda)^2$, and $K(m)$ and $E(m)$ are  the complete elliptic integrals of the
first and second kinds, respectively \cite{abramowitz1965}.
Their behaviour is shown in Fig. \ref{fig_Phi},
where we can see that $\Phi_2$ has its maximum value at $\lambda=0.98512$,
 and when $\lambda=1$, both
$\Phi_1$ and $\Phi_2$ vanish, despite the fact that $S_{\rm 1p}$ diverges at $k=0$.

From equations (\ref{eq11}), one can also show
\bea
S^{xx}_{\rm 1p}(k) & = & \frac{(1-\lambda^2)^{1/4}}{4\Lambda (k)}
\nonumber \\
S^{yy}_{\rm 1p}(k) & = & \frac{1}{4}(1-\lambda^2)^{1/4}\Lambda (k)
\label{xxyy}
\eea
Note that whereas $S^{xx}_{1p}(k)$ diverges as $\{\lambda \to 1, k=0\}$,
$S^{yy}_{1p}(k)$ does not.

By Fourier transforming, we obtain the 1-particle contributions to the
correlation functions (using equation (\ref{eq1a})):
\bea
C^{xx}_{\rm 1p}(\delta) & = &
(1-\lambda^2)^{1/4}\frac{1}{8\pi}\int^{2\pi}_0 dk
\frac{\cos(k\delta)}{\Lambda(k)}
\nonumber \\
C^{yy}_{\rm 1p}(\delta) & = &
(1-\lambda^2)^{1/4}\frac{1}{8\pi}\int^{2\pi}_0 dk
\cos(k\delta)\Lambda(k)
\label{Cxxyy}
\eea

Now exact expressions for the 1-particle contributions to the
correlation functions in the 2D
classical Ising model
have been obtained long ago by Wu, McCoy, Tracy and Barouch
\cite{wu1976,orrick2001}.
Taking the anisotropic `Hamiltonian limit' \cite{suzuki1976,fradkin1970} in which
the classical Ising model reduces to the quantum Ising model
(\ref{eq1}), one can show that the expressions of Wu {\it et al.} reduce
to equation (\ref{Cxxyy}) (details will be given elsewhere). This confirms
that the results we have given are in fact exact.

\begin{flushleft}
{\bf The Square Lattice}
\end{flushleft}

The transverse Ising model in (2+1) dimensions lies in the universality
class of the 3D classical Ising model, and so its critical exponents are
expected to be $\eta =0.0364(5)$ , $\nu =0.6301(4)$, from various estimates
\cite{pel02}.

The square lattice is bipartite, and hence one finds that the
Hamiltonian is symmetric under a spin rotation by $\pi$ about the $z$
axis on the B sublattice, followed by a coupling inversion $\lambda \to
-\lambda$. Correspondingly, there are symmetrical critical points at
couplings $\lambda = \pm \lambda_c$, where $\lambda_c$ has been
estimated from series expansions \cite{he1990} at 0.32851(8), and from a
finite-size scaling analysis \cite{hamer2000} at 0.32841(2).

We have computed series for the 1-particle spectral weight
$S^{xx}_{\rm 1p}({\bf k})$ and the quasiparticle energy $E({\bf k})$ to
order $\lambda^{14}$.
The calculation involved a list of 4\,654\,284  clusters, consisting of up to
15 sites.
 The leading
terms are
\begin{widetext}
\bea
E({\bf k}) & = & 2-2\lambda (\cos k_x +\cos k_y) +\lambda^2 [ 2-(\cos k_x
+ \cos k_y)^2] + \lambda^3 (\cos k_x + \cos k_y) [5-2(\cos k_x +
\cos k_y)^2]/2
\nonumber \\
 & &+ \lambda^4 [ 6+14(\cos k_x + \cos k_y)^2-5(\cos k_x + \cos
k_y)^4]/4 + O(\lambda^5)
\eea
\bea
S^{xx}_{1p}({\bf k}) & = &
1/4 + \lambda ( \cos k_x + \cos k_y )/4  +
  {\lambda}^2 [ 12\,\cos (k_x)\,\cos (k_y) +
      3 ( \cos (2\,k_x) + \cos (2\,k_y) ) ]/16  \nonumber \\
 &&  + \frac{ \lambda^3}{32}
       [15 (\cos k_x+ \cos k_y)
        + 30 (\cos (2k_x)\cos k_y + \cos k_x \cos (2k_y) ) + 5(\cos (3k_x)+ \cos (3k_y) )] + O(\lambda^4)
\eea
\begin{table}
\caption{Pole and residue of N/M Dlog Pad\'e approximants to $S_{\rm 1p}^{xx}$
for the square lattice. Defective approximants are marked with an
asterisk.}\label{tab2}
\begin{ruledtabular}
\begin{tabular}{|r|l|l|l|l|l|}
      &     N-2/N         &   N-1/N           &      N/N         &     N+1/N          &      N+2/N       \\
\hline
\multicolumn{6}{|c|}{$k=(0,0)$}\\
N= 1  &                  & 0.4000(-0.8000)  & 0.2703(-0.3652) &  0.3874(-1.0759)  & 0.2847(-0.3138) \\
N= 2  & 0.2953(-0.4682)  & 0.3263(-0.5900)  & 0.3285(-0.6022) &  0.3281(-0.5988)  & 0.3276(-0.5942) \\
N= 3  & 0.3286(-0.6028)  & 0.3281(-0.5996)  & 0.3319(-0.5789)*&  0.3278(-0.5969)  & 0.3282(-0.6017) \\
N= 4  & 0.3238(-0.5113)* & 0.3279(-0.5978)  & 0.3280(-0.5985) &  0.3288(-0.6122)  & 0.3284(-0.6037) \\
N= 5  & 0.3280(-0.5985)  & 0.3279(-0.5974)* & 0.3282(-0.6016) &  0.3283(-0.6035)  & 0.3284(-0.6036) \\
N= 6  & 0.3283(-0.6019)  & 0.3284(-0.6059)  & 0.3284(-0.6053) &  0.3284(-0.6050)  &                 \\
N= 7  & 0.3284(-0.6053)  & 0.3284(-0.6045)  &                 &                   &                 \\
\hline
\multicolumn{6}{|c|}{$k=(\pi/2,\pi/2)$}\\
N= 2  &                 & 0.6030(0.5455)  & 0.6030(0.5455) &  0.3428(0.0570)  & 0.3428(0.0570) \\
N= 3  & 0.6030(0.5455)  & 0.6030(0.5455)  & 0.3428(0.0570) &  0.3428(0.0570)  & 0.3586(0.0747) \\
N= 4  & 0.4231(0.1781)  & 0.3579(0.0735)  & 0.3579(0.0735) &  0.3521(0.0654)  & 0.3521(0.0654) \\
N= 5  & 0.3579(0.0735)  & 0.3579(0.0735)  & 0.3521(0.0654) &  0.3521(0.0654)  & 0.3425(0.0495) \\
N= 6  & 0.3473(0.0576)  & 0.3407(0.0459)  & 0.3407(0.0459) &  0.3390(0.0424)  &                \\
N= 7  & 0.3407(0.0459)  & 0.3407(0.0459)  &                &                  &                \\
\end{tabular}
\end{ruledtabular}
\end{table}
\end{widetext}
(further terms can be supplied on request). Note that because the
lattice is bipartite, both quantities are symmetric under the
simultaneous transformations $\{{\bf k} \to {\bf \pi - k}, \lambda \to -\lambda\}$.

The results of a standard Dlog Pad{\' e} analysis for $S^{xx}_{\rm 1p}({\bf k})$ at
${\bf k}=(0,0)$ and ${\bf k}=(\pi/2,\pi/2)$
are shown in Table \ref{tab2}.
At ${\bf k}=(0,0)$, where the energy gap vanishes, the table allows estimates  $\lambda_c= 0.3284(4)$ with
exponent $-0.605(5)$,
compared to the expected exponent $\nu(\eta-1) = -0.607$.
At momentum ${\bf k}=(\pi/2,\pi/2)$, where the energy gap remains finite,
we find $\lambda_c =0.34(3)$ with exponent $ 0.04(2)$ compared to the
expected value $\nu\eta = +0.0229$. In the latter case, the exponent
does not agree particularly well with the expected value,
but this inaccuracy might easily be attributed to the weakness of this
singularity. Complementary results follow for the antiferromagnetic
singularity at $\lambda = -0.3284$, with exponent $\nu(\eta-1)$ at ${\bf
k}=(\pi,\pi)$, and $\nu\eta$ at ${\bf k}=(\pi/2,\pi/2)$.

\begin{flushleft}
{\bf The Simple Cubic Lattice}
\end{flushleft}

The transverse Ising model in (3+1) dimensions lies in the universality
class of the 4D classical Ising model, where we expect the mean field exponents
 $\eta =0$ , $\nu =1/2$, modulo logarithmic corrections \cite{sachdev1999}.

We have computed series for the 1-particle
structure factor $S^{xx}_{\rm 1p}({\bf k})$
 and the quasiparticle
energy $E({\bf k})$  to orders $\lambda^{10}$.
The critical point has been obtained previously \cite{zheng1994} as
$\lambda_c = 0.19406(6)$.
A Dlog Pad{\' e} analysis of $S^{xx}_{\rm 1p}({\bf k})$ at ${\bf
k}=(0,0,0)$, where the energy gap vanishes, gives $\lambda_c= 0.19406(8)$ with
exponent $-0.54(1)$, while for another momentum, where the energy gap remains finite,
we find $\lambda_c =0.22(3)$ with exponent $ 0.03(2)$. Allowing for
logarithmic corrections, these agree reasonably well with the expected values.

In summary, then, we have
shown that the one-particle spectral weights of the transverse Ising
model satisfy the scaling behaviour at a quantum phase transition as
predicted by theory \cite{sachdev1999}. For the linear chain we obtain
new exact results; for the square and simple cubic lattices numerical
estimates are obtained from series expansions. Further details will be
published in a full length paper.

This work was supported by a grant from the Australian Research Council.
We are grateful for computational support from the Australian
Partnership for Advanced Computing (APAC) and the Australian Centre for
Advanced Computing and Communications (ac3).

\end{document}